\documentclass{aa}  

\usepackage{graphicx}
\usepackage{txfonts}
\begin{document}
\title{The Initial-Final Mass Relationship from white dwarfs in common 
       proper motion pairs$^\star$}
\author{S. Catal\'an
       \inst{1,2},
       J. Isern
       \inst{1,2},
       E. Garc\'\i a--Berro
       \inst{1,3},
       I. Ribas
       \inst{1,2},	  
       C. Allende Prieto
       \inst{4}
       \and 
       A.~Z. Bonanos
       \inst{5}
       }

\offprints{S. Catal\'an \\
$^\star$Based  on observations  obtained at:  Calar  Alto Observatory,
Almer\'\i a, Spain, el Roque  de los Muchachos, Canary Islands, Spain,
McDonald  Observatory,  Texas,  USA,  and  Las  Campanas  Observatory,
Chile.}

\institute{Institut d'Estudis Espacials de Catalunya, c/ Gran Capit\`{a} 
           2--4, 08034 Barcelona, Spain\
  	   \and	     
           Institut de Ci\`encies de l'Espai, CSIC, Facultat  de
           Ci\`encies, Campus UAB, 08193 Bellaterra, Spain\           
	   \and	     
	   Departament de F\'isica Aplicada, 
           Escola Polit\`ecnica Superior de Castelldefels, 
           Universitat Polit\`ecnica de Catalunya, 
           Avda. del Canal Ol\'\i mpic s/n, 08860 
	   Castelldefels, Spain
	   \and
	   McDonald Observatory and Department of Astronomy, 
           University of Texas, Austin, TX 78712, USA
	   \and
	   Vera Rubin Fellow, Department of Terrestrial Magnetism, 
           Carnegie Institution of Washington, DC 20015, USA }
\date{\today}

\abstract 
{The initial-final mass relationship of white dwarfs, which is poorly constrained, is of paramount 
importance for different aspects in modern astrophysics. From an observational perspective, most of 
the studies up to now have been done using white dwarfs in open clusters.}
{In order to improve the initial-final mass relationship we explore the possibility of deriving a 
semi-empirical relation studying white dwarfs in common proper motion pairs. If these systems are 
comprised of a white dwarf and a FGK star, the total age and the metallicity of the progenitor of 
the white dwarf can be inferred from the detailed analysis of the companion.}
{We have performed an exhaustive search of common proper motion pairs containing a DA white dwarf 
and a FGK star using the available literature and crossing the SIMBAD database with the Villanova 
White Dwarf Catalog.  We have acquired long-slit spectra of the white dwarf members of the selected 
common proper motion pairs, as well as high resolution spectra of their companions.  From these 
observations, a full analysis of the two members of each common proper motion pair leads to the 
initial and final masses of the white dwarfs.}
{These observations have allowed us to provide updated information for the white dwarfs, since some 
of them were misclassified.  In the case of the DA white dwarfs, their atmospheric parameters, 
masses, and cooling times, have been derived using appropriate white dwarf models and cooling 
sequences.  From a detailed analysis of the FGK stars spectra we have inferred the metallicity. Then, 
using either isochrones or X-ray luminosities we have obtained the main-sequence lifetime of the 
progenitors, and subsequently their initial masses.}
{This work is the first one in using common proper motion pairs to improve the initial-final mass relationship, and has also allowed to cover the poorly explored low-mass domain. As in the case of studies based on white 
dwarfs in open clusters, the distribution of the semi-empirical data presents a large scatter, 
which is higher than the expected uncertainties in the derived values.  This suggests that the 
initial-final mass relationship may not be a single-valued function.}
\authorrunning{Catal\'an et al.}
\titlerunning{The IFMR from WDs in common proper motion pairs}

\maketitle


\section{Introduction}

White dwarfs are the final remnants of low- and intermediate-mass stars. About 95\% of 
main-sequence stars will end their evolutionary pathways as white dwarfs and, hence, the study of 
the white dwarf population provides details about the late stages of the life of the vast majority 
of stars.  Since white dwarfs are long-lived objects, they also constitute useful objects to study 
the structure and evolution of our Galaxy (Liebert et al.~2005a; Isern et al. 2001). For instance, 
the initial-final mass relationship (IFMR), which connects the properties of a white dwarf with 
those of its main-sequence progenitor, is of paramount importance for different aspects in modern 
astrophysics. It is required as an input for determining the ages of globular clusters and their 
distances, for studying the chemical evolution of galaxies, and also to understand the properties 
of the Galactic population of white dwarfs.  Despite its relevance, this relationship is still 
poorly constrained, both from the theoretical and the observational points of view.

The first attempt to empirically determine the initial-final mass relationship was undertaken by 
\cite{wei77}, who also provides a recent review on this subject (Weidemann 2000). It is still not 
clear how this function depends on the mass and metallicity of the progenitor, its angular 
momentum, or the presence of a strong magnetic field. The total age of a white dwarf can be 
expressed as the sum of its cooling time and the main-sequence lifetime of its progenitor. The 
latter depends on the metallicity of the progenitor of the white dwarf, but it cannot be determined 
from observations of single white dwarfs.  This is because white dwarfs have such strong surface 
gravities that gravitational settling operates very efficiently in their atmospheres, and any 
information about their progenitors (e.g. metallicity)  is lost in the very early evolutionary 
stages of the cooling track. Moreover, the evolution during the AGB phase of the progenitors is 
essential in determining the size and composition of the atmospheres of the resulting white dwarfs, 
since the burning processes that take place in H and He shells determine their respective 
thicknesses and their detailed chemical compositions, which are crucial ingredients for determining 
the evolutionary cooling times.

A promising approach to circumvent the problem, and also to directly test the initial-final mass 
relationship, is to study white dwarfs for which external constraints are available.  This is the 
case of white dwarfs in open and globular clusters (Ferrario et al.~2005, Dobbie et al.~2006) or in 
non-interacting binaries, for instance, common proper motion pairs (Wegner 1973, Oswalt et 
al.~1988).  Focusing on the latter, it is sound to assume that the members of a common proper 
motion pair were born simultaneously and with the same chemical composition. Since the components 
are well separated (100 to 1000 AU), mass exchange between them is unlikely and it can be 
considered that they have evolved as isolated stars. Thus, important information of the white 
dwarf, such as its total age or the metallicity of the progenitor, can be inferred from the study 
of the companion.  In particular, if the companion is an F, G or K type star the metallicity can be 
derived with high accuracy from detailed spectral analysis. On the other hand, the age can be 
obtained using different methods. In particular, we will use stellar isochrones when the star is 
moderately evolved, or the X-ray luminosity if the star is very close to the ZAMS.

The purpose of this work is to present our spectroscopic analysis of both members of some common 
proper motion pairs containing a white dwarf, and the semi-empirical intial-final mass relationship 
that we have derived from this study. The paper is organized as follows.  In \S 2 we present the 
observations done so far and describe the data reduction.  Section 3 is devoted to discuss the 
classification and the analysis of the observed white dwarfs, whereas in \S 4 we present the 
analysis of the companions. This is followed by \S 5 where we present our main results and finally 
in \S 6 we elaborate our conclusions.


\section{Observations and data reduction}

The sample of common proper motion pairs to be observed was chosen from the available literature, 
mainly from the papers of \cite{sil01} and \cite{weg91}, and from a cross-correlation of the SIMBAD 
database and the Villanova White Dwarf Catalog.  We selected the pairs taking into account 
different requirements.  Firstly, the white dwarf component should be classified as a DA (i.e., 
with the unique presence of Balmer lines), so that the fitting procedure is sufficiently accurate 
to derive realistic values for the effective temperature and surface gravity.  Secondly, the other 
component of the pair should be a star of spectral type F, G or K for an accurate determination of 
the metallicity, and moderately evolved or very close to the ZAMS in order to be able to estimate 
its age.  The complete list of targets is given in Table \ref{tab:1}.

The observations were carried out during different campaigns between the summer of 2005 and the 
spring of 2007.  In Table \ref{tab:2} we give details of the telescope-instrument configurations 
employed, as well as the resolution and spectral coverage of each setup.

For the white dwarf members we performed long-slit low-resolution spectroscopic observations 
covering some of the main Balmer lines (from H$\beta$ to H$8$).  WD0315$-$011 was kindly observed 
for us by T.~Oswalt with the RC spectrograph at the 4~m telescope at Kitt Peak National Observatory 
with a resoltuion of about $1.5$~\AA~ FWHM.  We performed as many exposures as necessary to 
guarantee a high signal-to-noise ratio final spectrum for each object (after the corresponding 
reduction).  Spectra of high quality are essential to derive the atmospheric parameters with 
accuracy.  We co-added individual 1800 s~ exposures to minimize the effects of cosmic ray impacts 
on the CCD.

The white dwarf spectra were reduced using the standard procedures within the single-slit tasks in 
IRAF\footnote{IRAF is distributed by the National Optical Astronomy observatories, which are 
operated by the Association of Universitites for Research in Astronomy, Inc., under cooperative 
agreement with the national Science Foundation ({\tt http://iraf.noao.edu}).}.  First, the images 
were bias- and flatfield-corrected, and then, the spectra were extracted and wavelength calibrated 
using arc lamp observations.  We combined multiple spectra of the same star to achieve a final 
spectrum of high signal-to-noise ratio (S/N $>$ 100).  Before this step, we applied the 
heliocentric correction of each spectrum, since we were co-adding spectra secured in different 
days. Finally, they were normalized to the continuum.

The FGK companions were observed with echelle spectrographs, obtaining high signal-to-noise 
high-resolution spectra (S/N $>$ 150), which are necessary to derive the metallicity with accuracy.  
For the reduction of the FGK stars spectra the procedure followed was similar to the case of white 
dwarfs but we used the corresponding echelle tasks in IRAF. In this case, we used the task 
{\tt apscatter} in order to model and subtract the scattered light.

\begin{table}
\begin{center}
\caption{Common proper motion pairs studied in this work.}
\begin{tabular}{lcll}
\hline
\hline
\noalign{\smallskip}  
System  &  White   Dwarf  &  Companion & Sp. Type$^1$\\
 & & \\
\noalign{\smallskip}
\hline
\noalign{\smallskip}      
G 158$-$78/77	  & WD0023$-$109 & G158$-$77     & K    \\ 
LP 592$-$80/      & WD0315$-$011 & BD~$-$01~469A & K1IV \\
LTT 1560          &              &               &      \\
GJ 166 A/B	  & WD0413$-$077 & HD~26965      & K1V  \\
G 116$-$16/14     & WD0913$+$442 & BD~$+$44~1847 & G0   \\  
G 163$-$B9B/A     & WD1043$-$034 & G163$-$B9A    & F9V  \\
LP 378$-$537      & WD1304$+$227 & BD~$+$23~2539 & K0   \\ 
G 165$-$B5B/A     & WD1354$+$340 & BD~$+$34~2473 & F8   \\ 
G 66$-$36/35      & WD1449$+$003 & G66$-$35      & G5V  \\ 
EGGR 113/         & WD1544$+$008 & BD~$+$01~3129 & G0   \\   
BD~$+$01~3129     &              &               &      \\   
GJ 599 A/B	  & WD1544$-$377 & HD~140901     & G6V  \\ 
GJ 620.1 B/A      & WD1620$-$391 & HD~147513     & G5V  \\
GJ 2125 / GJ 3985 & WD1659$-$531 & HD~153580     & F6V  \\
G 140$-$B1B/      & WD1750$+$098 & BD~$+$09~3501 & K0   \\ 
BD~$+$09~3501 	  &              &               &      \\ 
G 156$-$64/65     & WD2253$-$081 & BD~$-$08~5980 & G6V  \\ 
\noalign{\smallskip}
\hline
\hline 
\end{tabular}
\label{tab:1}
\end{center}
\small \footnotemark[1]{From SIMBAD database.}\\
\end{table}

\begin{table*}
\begin{center}
\caption{Journal of observations.}
\begin{tabular}{lccccc}
\hline
\hline
\noalign{\smallskip} 
Observatory  & Telescope & Spectrograph & R & Spectral \\
  &   &  &  & Coverage \\
\noalign{\smallskip}
\hline
\noalign{\smallskip}  
& & White Dwarfs & & \\
\hline
McDonald     & 2.7~m HJS      & LCS       &  1,000 & 3885$-$5267~\AA  \\ 
CAHA         & 3.5~m          & TWIN      &  1,250 & 3570$-$5750~\AA  \\ 
LCO          & 6.5~m Clay     & LDSS3     &  1,650 & 3600$-$6000~\AA  \\  
\hline
& & Low-mass Companions & & \\
\hline
McDonald     & 2.7~m HJS      & 2dcoud\'e & 60,000 & 3400$-$10900~\AA \\ 
CAHA         & 2.2~m          & FOCES     & 47,000 & 3600$-$9400~\AA  \\ 
ORM          & 3.5~m TNG      & SARG      & 57,000 & 4960$-$10110~\AA \\ 
LCO          & 6.5~m Clay     & MIKE      & 65,000 & 4900$-$10000~\AA \\  
\noalign{\smallskip}
\hline
\hline
\end{tabular}
\label{tab:2}
\end{center}
\end{table*}

\section{White dwarf analysis}
\subsection{Classification}

\begin{table*}[htbp]
\begin{center}
\caption{Spectral classification of the  white  dwarfs.}
\begin{tabular}{lclc}
\hline
\hline
\noalign{\smallskip}
Name & This Work & Previous  & Reference \\
\noalign{\smallskip}
\hline
\noalign{\smallskip}
WD0023$-$109 & DA      & DA     & EG65, WR91, OS94, MS99 \\
WD0315$-$011 & DA      & DA     & OS94, MS99, SOW01 \\
WD0413$-$077 & DA      & DA     & EG65, FKB97, MS99, HOS02, HBB03, KNH05, HB06 \\ 
WD0913$+$442 & DA      & DA     & EG65, WR91, BLF95, BLR01, ZKR03, KNH05, LBH05, HB06 \\
WD1043$-$034 & sdB$^1$ & DA/sd  & WR91 \\
             &         & DAB    & OS94, MS99 \\
WD1304$+$227 & DA      & DA     & O81, OS94, MS99, SOW01 \\
WD1354$+$340 & DA      & DA     & EG67, WR91, BLF95, MS99, SOW01 \\
WD1449$+$003 & M       & DA     & O81, WR91, OS94 \\
             &         & M      & FBZ05 \\
WD1544$+$008 & sdO$^1$ & DA/sdO & W91 \\
	     &         & DA     & EG65, MS99 \\
	     &         & DAB    & SOW01 \\
WD1544$-$377 & DA      & DA     & EG65, W73, OS94, PSH98, BLR01, KNC01, SOW01 \\ 
             &         & DA     & HOS02, HBB03, ZKR03, KNH05, KVS07 \\ 
WD1620$-$391 & DA      & DA     & W73, OS94, HBS98, PSH98, SOW01, HOS02, HBB03, HB06, KVS07 \\
WD1659$-$531 & DA      & DA     & W73, OS94, PSH98, SOW01, KVS07 \\
WD1750$+$098 & DC      & DA     & WR91, SOW01 \\
             &         & DC     & EG65, OS94, MS99 \\
WD2253$-$081 & DA      & DA     & OS94, BLF95, MS99, BLR01, KNC01, SOW01, KNH05 \\
\noalign{\smallskip}
\hline
\hline
\end{tabular}
\label{tab:3}
\smallskip 
\end{center}
\small \footnotemark[1]{P.~Bergeron, private communication.}\\
\\
References.~(BLF95) \cite{ber95a}; (BLR01) \cite{ber01a}; (EG65)  \cite{egg65}; 
            (EG67)  \cite{egg67};  (FBZ05) \cite{far05a}; (FKB97) \cite{fin97};
            (HOS02) \cite{hol02};  (HBB03) \cite{hol03};  (HB06)  \cite{hol06}; 
            (KNC01) \cite{koe01};  (KNH05) \cite{kar05a}; (KVS07) \cite{kaw07a};
            (MS99)  \cite{mcc99a}; (O81)   \cite{osw81};  (OS94)  \cite{osw94}; 
            (PSH98) \cite{pro98};  (SOW01) \cite{sil01a}; (W73)   \cite{weg73a};
            (WR91) \cite{weg91a};  (ZKR03) \cite{zuc03}
\end{table*}

After the corresponding reduction, we carried out a first inspection of the spectra.  All the 
objects in Table \ref{tab:3} were previously classified as DA white dwarfs.  However, we found that 
four of them are not of DA type. Particularly, WD1750$+$098 turned out to be of type DC although in 
the most recent reference (Silvestri et al.~2001) it appears classified as a DA. We believe that 
WD1544$+$008 is the same star as WD1544$+$009, which was classified as a DAB white dwarf by 
\cite{sil01}.  However, it was identified as a sdO star by \cite{weg91}.  The same authors also 
studied WD1043$-$034 and classified it as a sdB star, although \cite{mcc99} considered it as a DAB 
white dwarf.  Taking into account the different inconsistencies in the literature, we decided to 
reobserve these objects in order to revise their spectral classifications, if necessary.  The 
reduced spectra of these two stars were kindly analysed by P.~Bergeron, who performed the 
corresponding fits and derived their temperatures and surface gravities, which turned out to be too 
low to be white dwarfs.  As can be seen in Table \ref{tab:3}, WD1449$+$003 is an M star.  This 
classification was also recently indicated by \cite{far05}.  These authors also reported that 
WD0913$+$442 and BD~$+$44~1847 are not a physical pair according to their parallaxes. It is worth 
mentioning that some of the previous misclassifications are probably due to the fact that the 
signal-to-noise ratio of the spectra used was low.

\subsection{Atmospheric parameters}

\begin{table}
\begin{center}
\caption{Atmospheric parameters derived for the observed white dwarfs.}
\label{tab:4}
\begin{tabular}{lcc}
\hline
\hline
\noalign{\smallskip}
Name & $T_{\rm eff}$~(K) & $\log g$ (dex) \\
\noalign{\smallskip}
\hline
\noalign{\smallskip}
WD0023$+$109     &$10380\pm230$ &$7.92\pm0.08$ \\
WD0315$-$011     &$7520\pm260$  &$8.01\pm0.45$ \\
WD0413$-$077$^1$ &$16570\pm350$ &$7.86\pm0.05$ \\
WD0913$+$442     &$8920\pm110$  &$8.29\pm0.10$ \\
WD1304$+$227     &$10800\pm120$ &$8.21\pm0.05$ \\
WD1354$+$340     &$13650\pm420$ &$7.80\pm0.15$ \\
WD1544$-$377     &$10600\pm250$ &$8.29\pm0.05$ \\
WD1620$-$391     &$24900\pm130$ &$7.99\pm0.03$ \\
WD1659$-$531     &$14510\pm250$ &$8.08\pm0.03$ \\
WD2253$-$081     &$7220\pm140$  &$8.25\pm0.20$ \\
\noalign{\smallskip}
\hline
\hline
\end{tabular}
\end{center}
\small \footnotemark[1]{We do not have  a spectrum of this star. These
                        values     are    from     \cite{heb97}    and
                        \cite{ber95}.}\\
\end{table}

Before calculating the atmospheric parameters of the white dwarfs ($T_{\rm eff}$ and $\log g$)  we 
determined the radial velocities of each star using the IRAF task {\tt fxcor}. Each spectrum was 
cross-correlated with a reference model from a grid computed by D.~Koester (private communication). 
The obtained radial velocities, estimated with large error bars, were generally small (ranging from 
10 to 50~km/s)  compared with the resolution element (300~km/s) of our observations.  In only one 
case (WD0023$+$109), the radial velocity measured turned out to be relevant (150~km/s). However, 
all radial velocities were taken into account for consistency.

After this previous step, we derived the atmospheric parameters of these stars performing a fit of 
the observed Balmer lines to white dwarf models following the procedure described in \cite{ber92}. 
The models had been previously normalized to the continuum and convolved with a Gaussian 
instrumental profile with the proper FWHM in order to have the same resolution as the observed 
spectra. The fit of the line profiles was then carried out using the task {\tt specfit} of the IRAF 
package, which is based on $\chi^2$ minimization with the Levenberg-Marquardt method. We used {\tt 
specfit} for different $\log g$ values (7.0, 7.5, 8.0, 8.5 and 9.0) with $T_{\rm eff}$ as a free 
parameter, obtaining different $\chi^2$ for each fit.  In each case, the initial estimate for 
$T_{\rm eff}$ obtained from the spectral energy distribution (photometry in the $BV$ and $JHK$ 
bands, 2MASS) was used as a starting guess.  The uncertainties in the derived $T_{\rm eff}$ were 
estimated from the perturbations required to increase the value of the reduced $\chi^2$ by one.

The determination of $\log g$ was performed in an analogous way but to calculate the errors we took 
into account the prescription of \cite{ber92}, who derive them from the independent fits of the 
individual exposures for any given star (before the combination).  The results are given in Table 
\ref{tab:4}.  In Fig.~1 we show the fits for some of the DA white dwarfs in our sample.

\begin{figure*}[t]
\begin{center}
\includegraphics[scale=0.4]{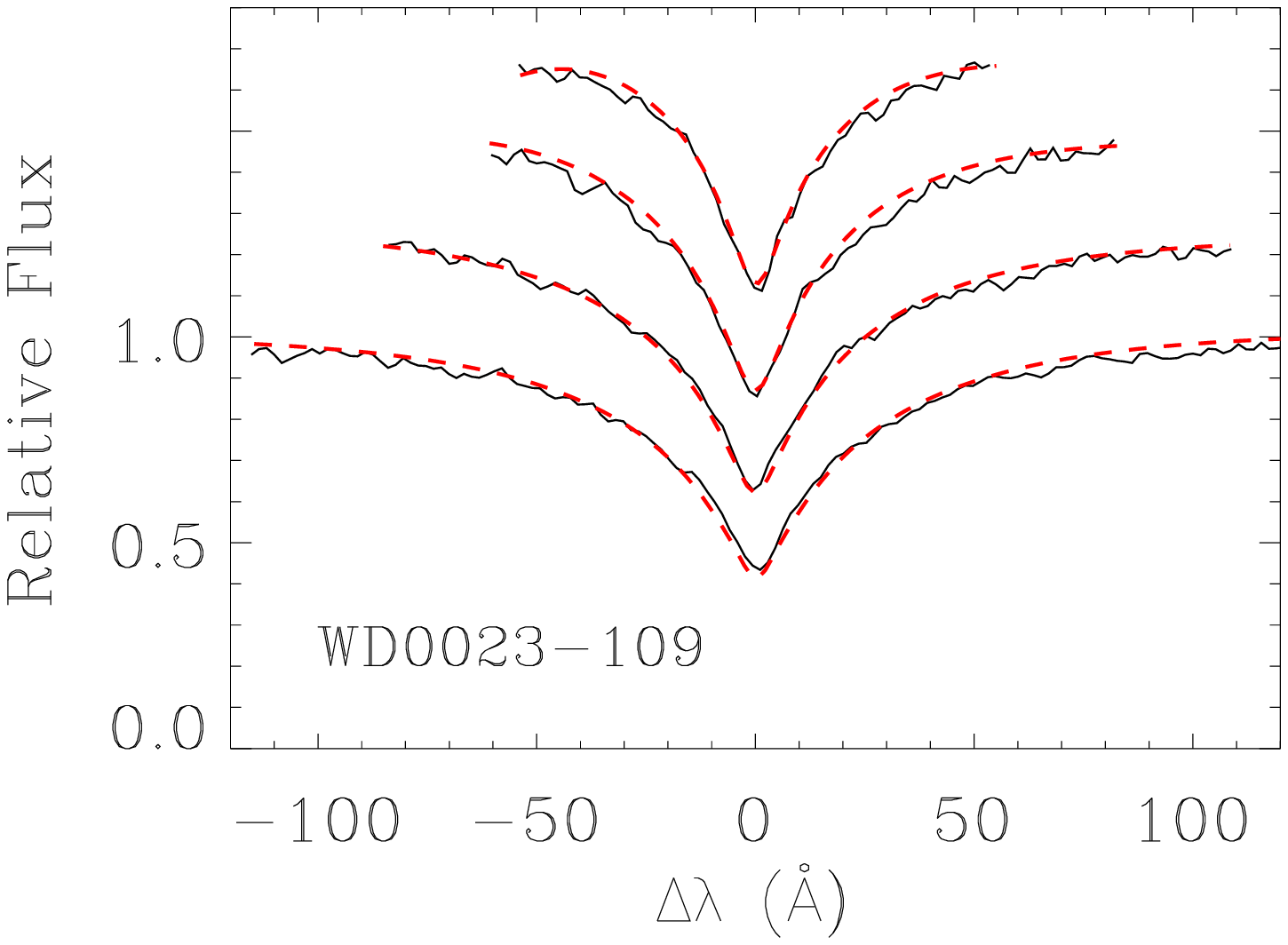}
\includegraphics[scale=0.4]{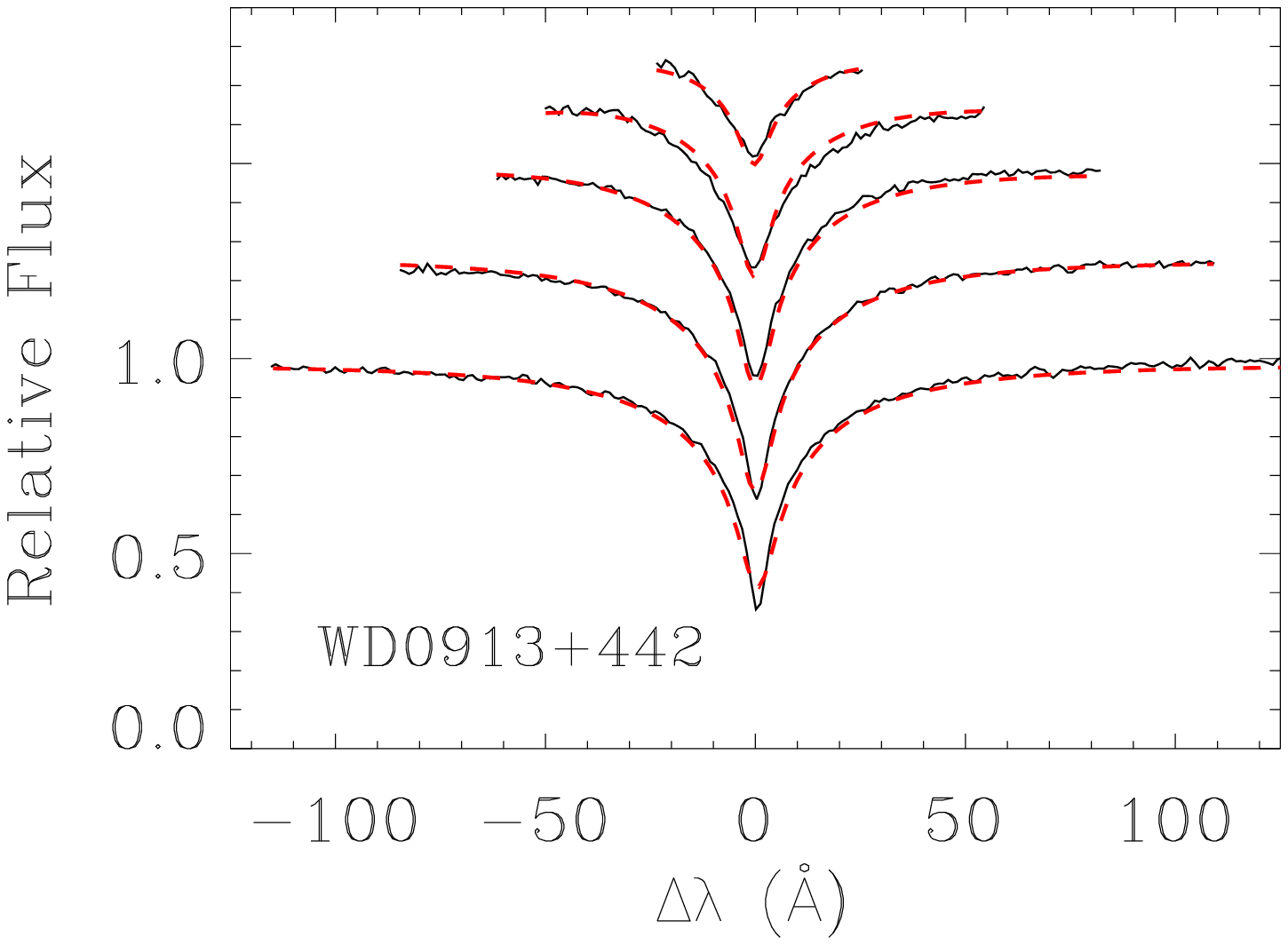}
\includegraphics[scale=0.4]{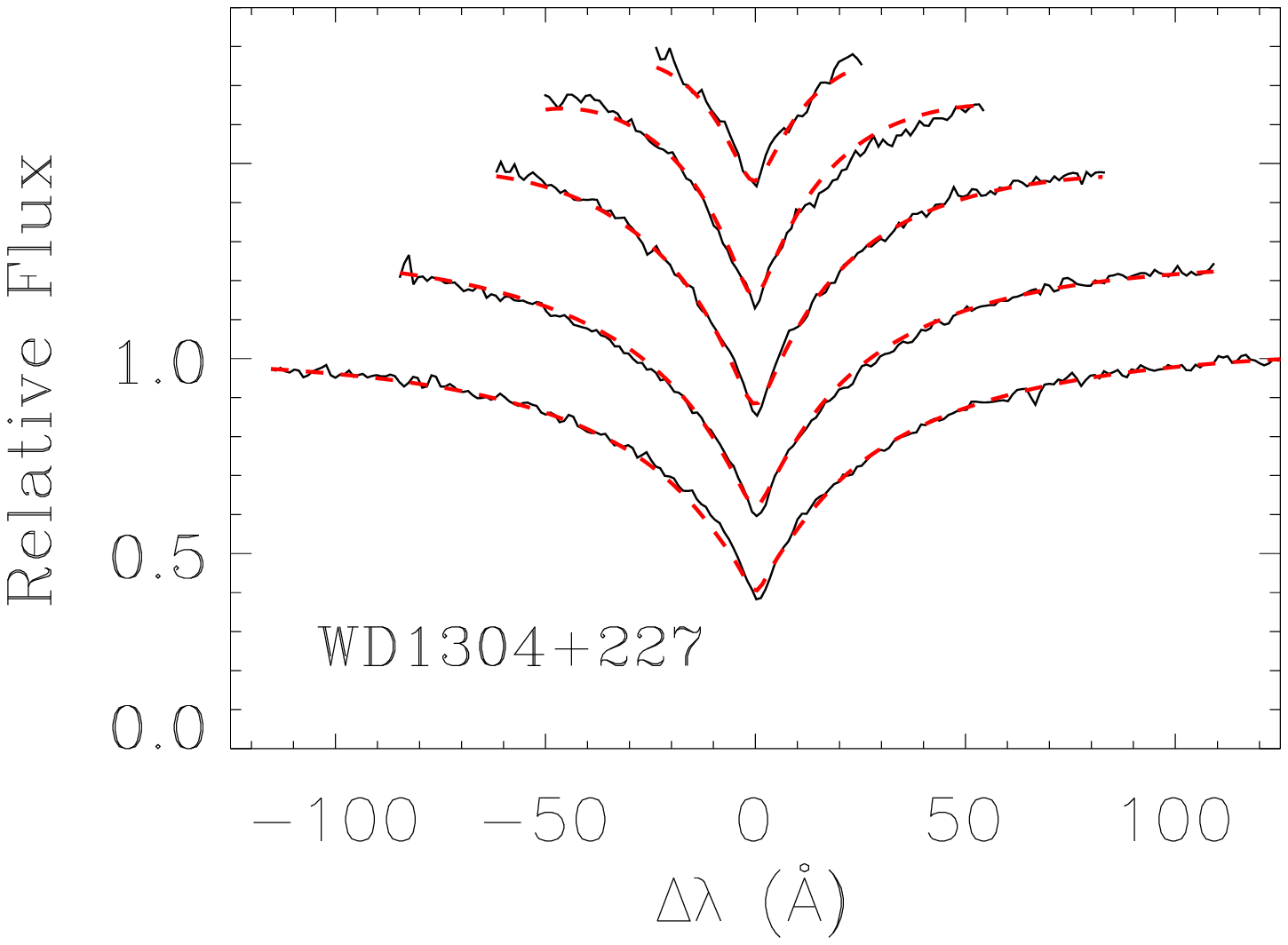}
\includegraphics[scale=0.4]{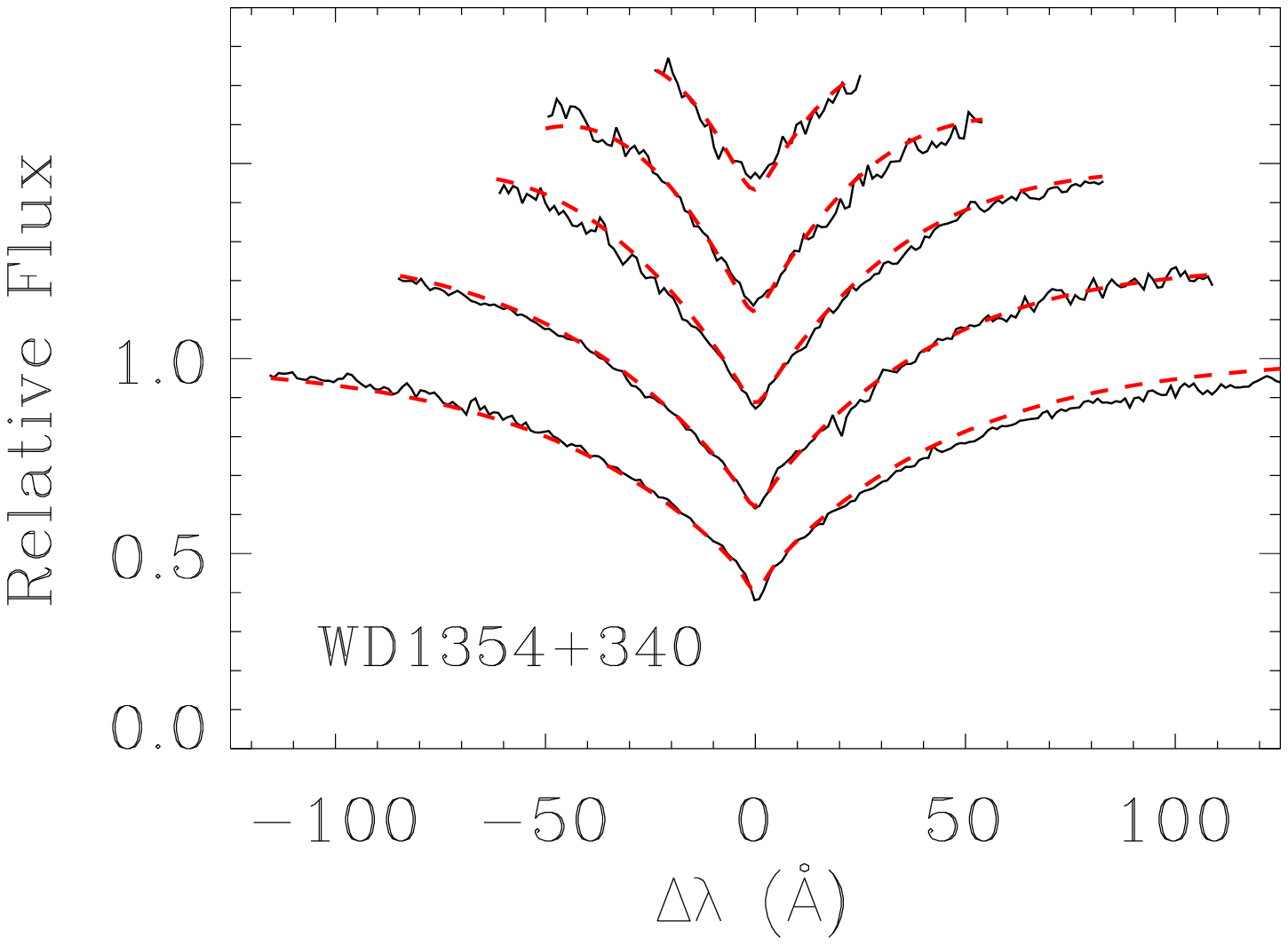}
\includegraphics[scale=0.4]{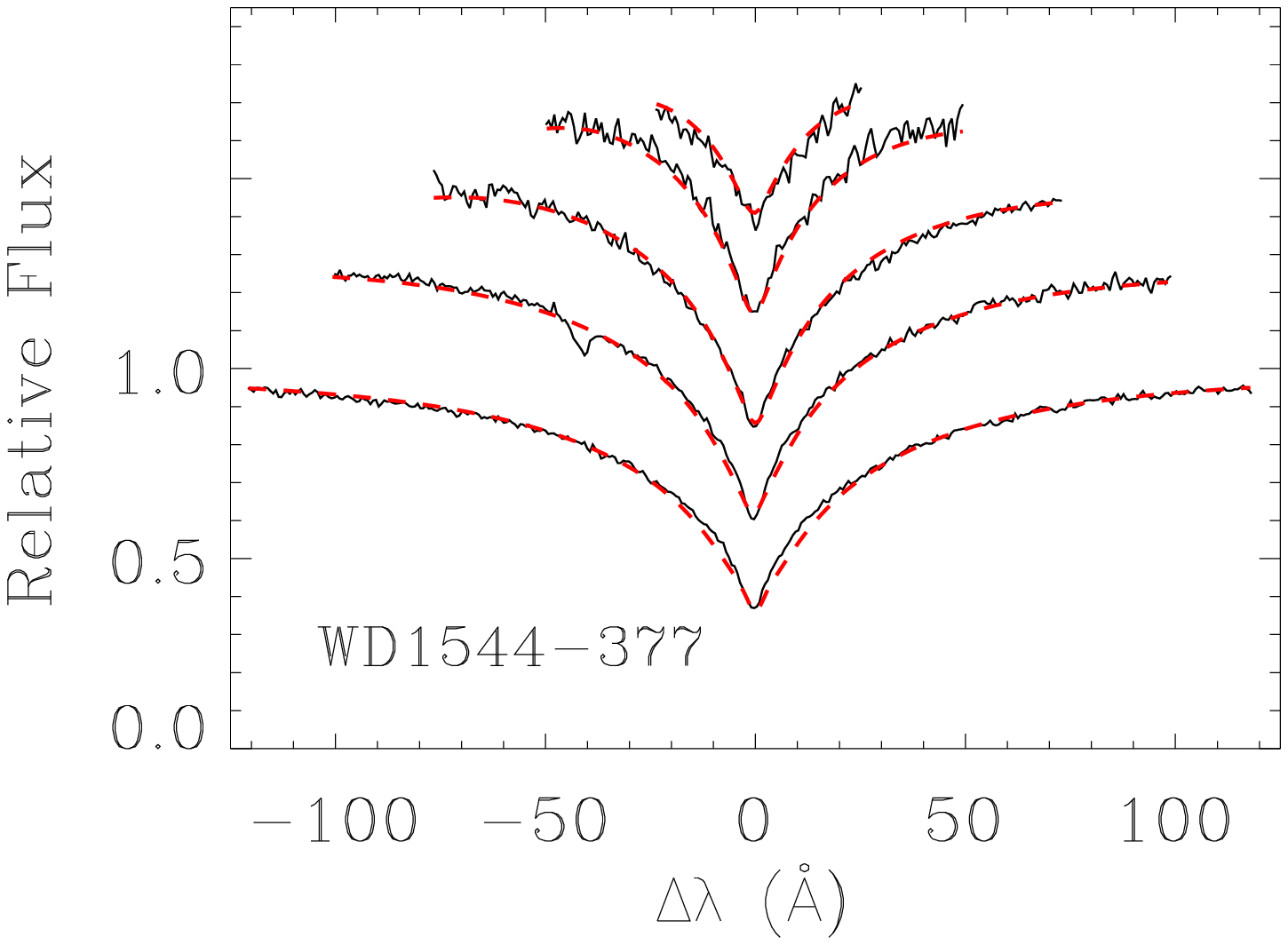}
\includegraphics[scale=0.4]{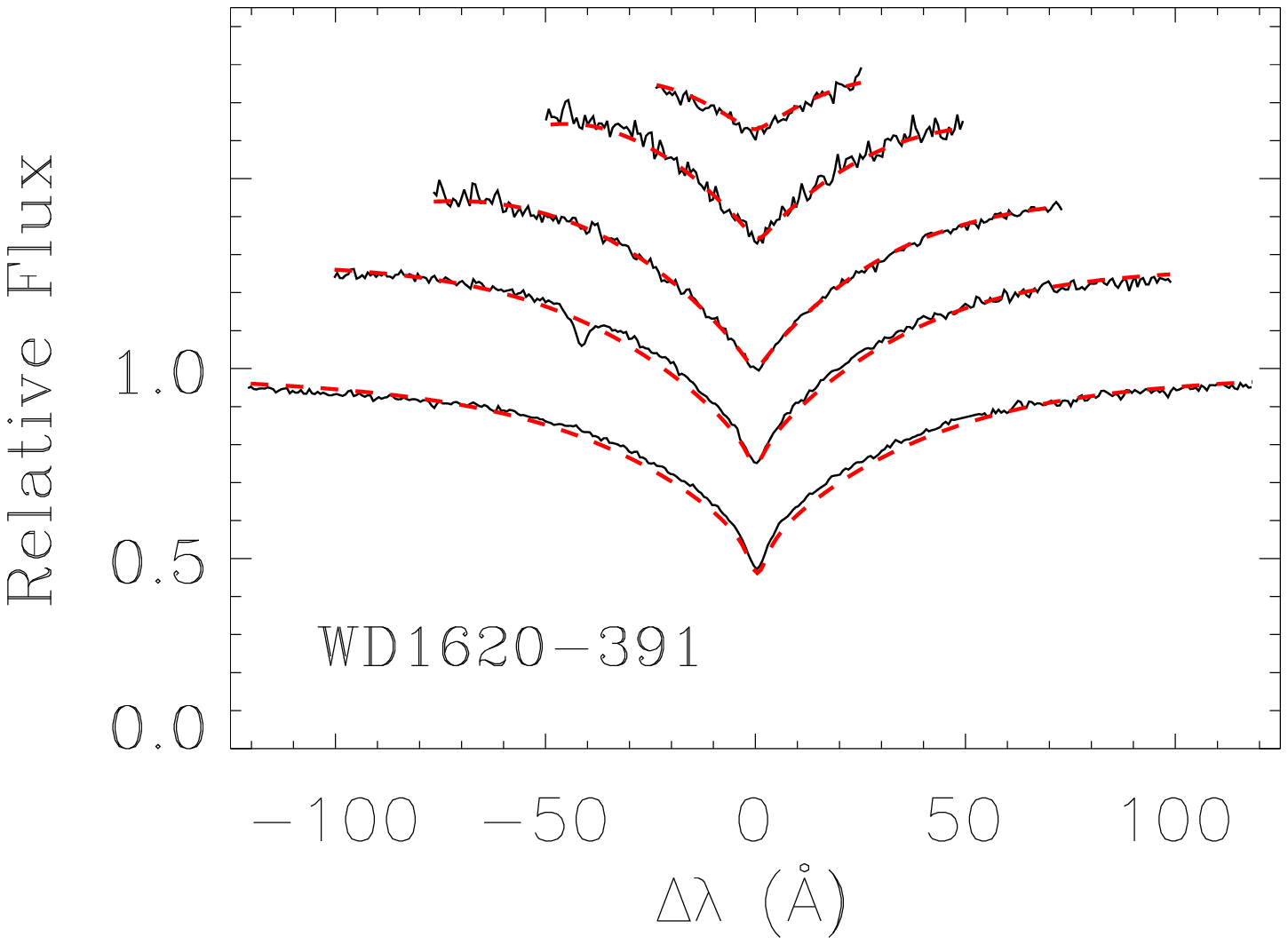}
\includegraphics[scale=0.4]{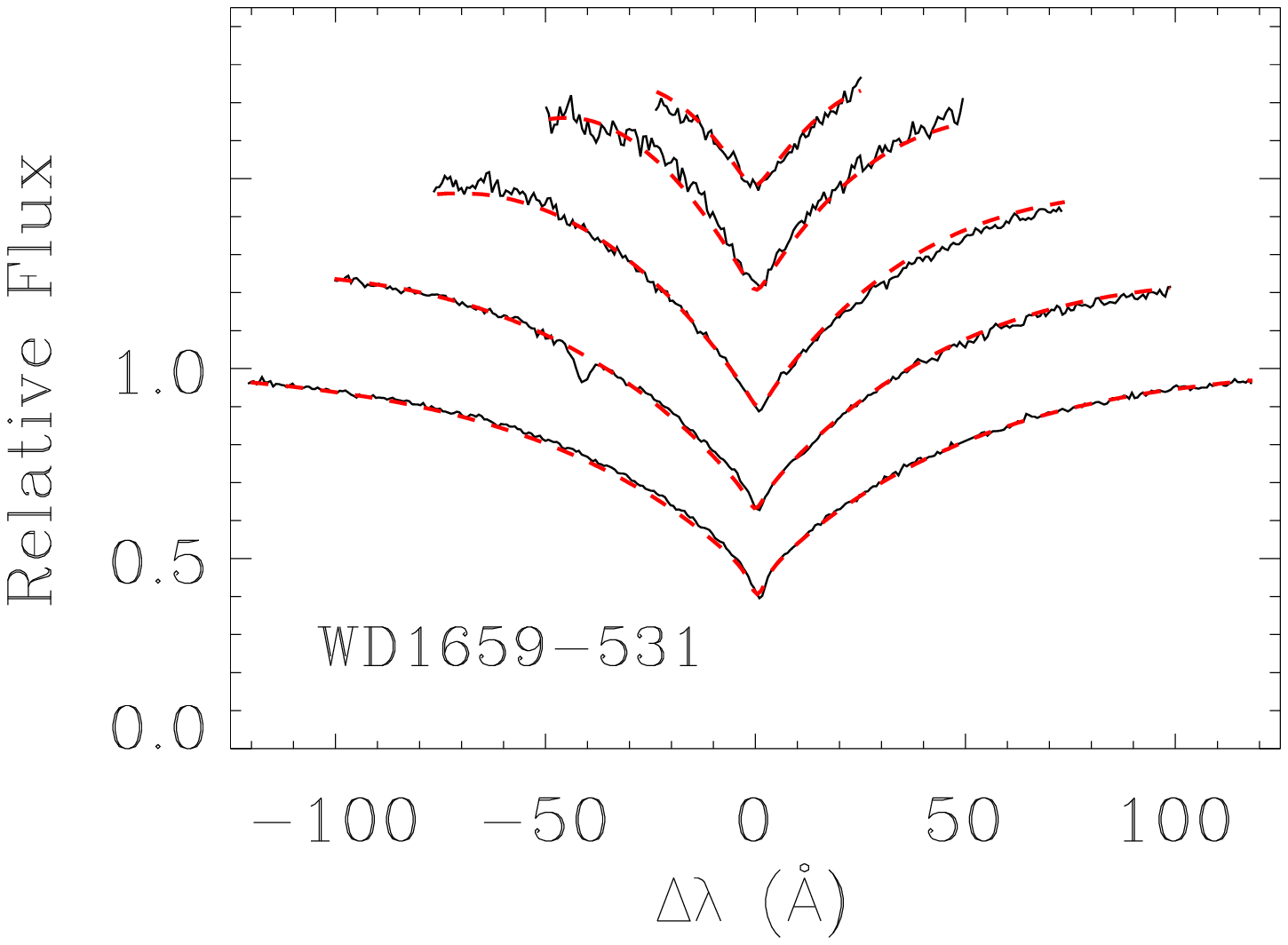}
\includegraphics[scale=0.4]{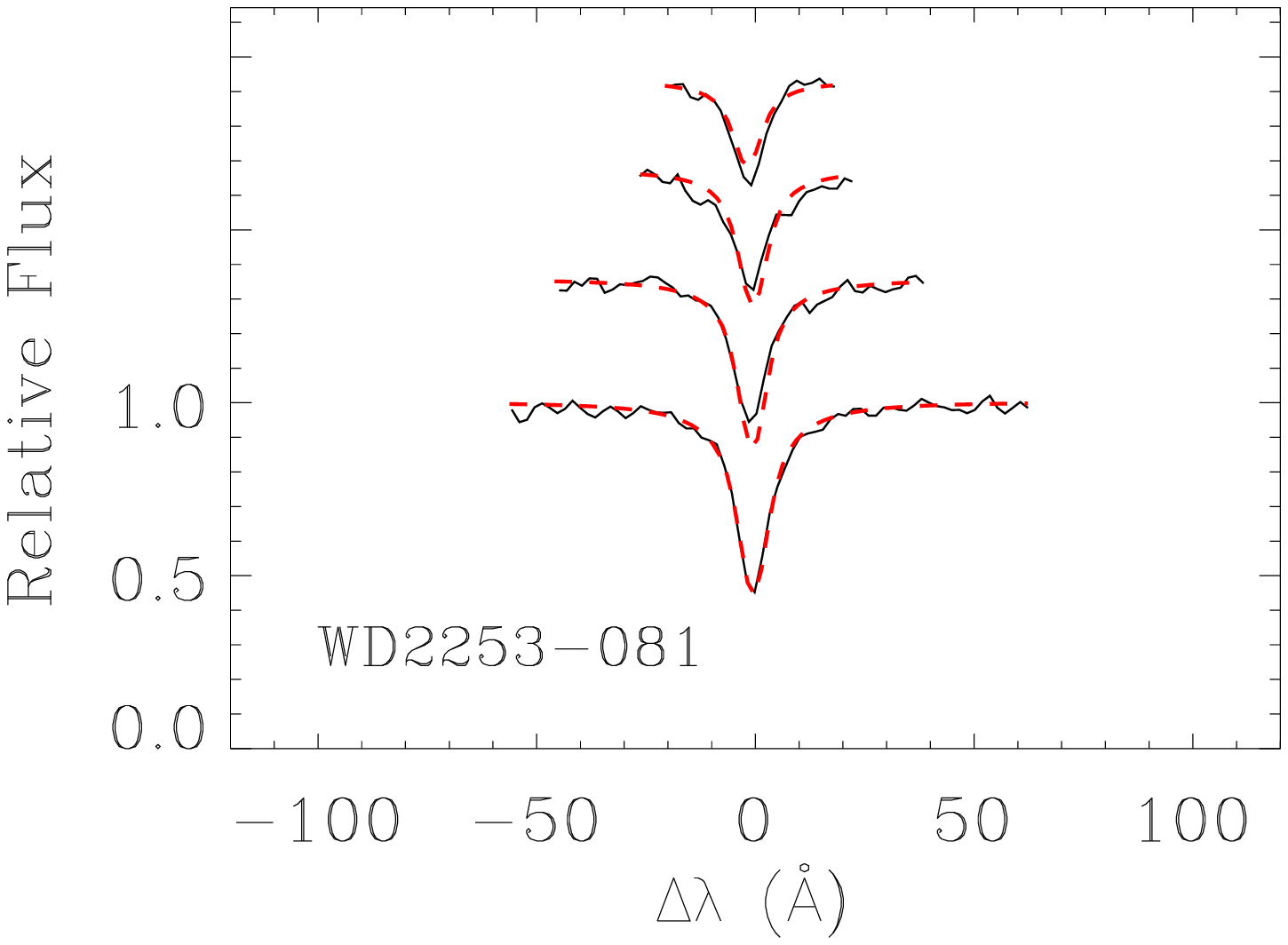}
\caption{Fits  of  the observed  Balmer  lines  for  the white  dwarfs
         studied  here.  Lines  range from  H$\beta$ (bottom)  to H$8$
         (top),  except   for  WD0023$-$109   and   WD2253$-$081   (to
         H$\varepsilon$).}
\end{center}
\label{fig:fitswd}
\end{figure*}

Some of these white dwarfs had been the subject of previous analyses which allow us to perform a 
comparison with our results.  For instance, WD0913$+$442 was also studied by \cite{ber95}, who 
obtained atmospheric parameters compatible with the ones derived here.  They also studied 
WD1354$+$340 and WD2253$-$081, but in these cases the effective temperatures obtained are 
compatible with ours while the surface gravities are not, although just outside the $1\sigma$ error 
bar. We have obtained lower values of $\log g$ in both cases, which could be due to the different 
resolution of the spectra ($\sim6$ FHWM in their case). This latter object, WD2253$-$081, is of 
particular interest since an accurate fit of its line profiles posed many problems to previous 
analyses because the lines seemed to be broader than the models predicted.  This led different 
authors to consider the possibility of this star to be a magnetic white dwarf or to have its lines 
rotationally broadened.  Both options were considered by \cite{kar05}, who discarded the former 
possibility.  With the purpose of solving the fitting problem of this star, in this work we have 
used updated models for DA white dwarfs with effective temperatures between 6000 and 10000 K.  
These models were kindly provided by D.~Koester, who calculated them considering collision-induced 
absorption due to the presence of molecular hydrogen.  This effect is very significant at low 
temperatures and it should be taken into account for an accurate determination of the atmospheric 
parameters. Contrarily to the results obtained by \cite{kar05} we did not need to consider 
rotational broadening to achieve a good fit. On the other hand, the southern hemisphere targets had 
been also studied by different authors.  Recently, \cite{kaw07} derived the atmospheric parameters 
for WD1544$-$377, WD1620$-$391 and WD1659$-$531, which are in good agreement with our results.

\subsection{Masses and cooling times}

Once we have derived the $T_{\rm eff}$ and $\log g$ of each star, we can obtain its mass ($M_{\rm 
WD}$) and cooling time ($t_{\rm cool}$) from appropriate cooling sequences.  We have used the 
cooling tracks of \cite{sal00} --- model S0 --- which consider a carbon-oxygen (C/O) core white 
dwarf (with a higher abundance of O at the center of the core) with a thick hydrogen envelope on 
top of a helium buffer, $q({\rm H})=M_{\rm H}/M=10^{-4}$ and $q({\rm He})=M_{\rm He}/M=10^{-2}$. 
These improved cooling sequences include an accurate treatment of the crystallization process of 
the C/O core, including phase separation upon crystallization, together with up-to-date input 
physics suitable for computing white dwarf evolution. In order to check the sensitivity of our 
results to the adopted cooling tracks, we also used the sequences of \cite{fon01} with different 
core compositions.  In a first series of calculations, C/O cores with a composition of 50/50 by 
mass with thick H envelopes, $q({\rm H})=10^{-4}$, on top of a He buffer, $q({\rm He})=10^{-2}$, 
were adopted.  We refer to these models as F0. In the second series of calculations, cooling 
sequences with a pure C core and the same envelope characteristics --- model F1 --- were used.  As 
can be seen in Table \ref{tab:5}, the derived masses do not change appreciably when adopting 
different cooling sequences. On the contrary, small differences can be noted in the cooling times 
obtained, depending on the evolutionary tracks used.  This stems naturally from the different core 
compositions of the cooling sequences adopted here.  As can be noted by examining Table 5, 
considering a C/O core with equal carbon-oxygen mass fractions with thick envelopes (model F0) is 
quite similar to considering a C/O core with more O concentrated in the center of the core (model 
S0) in terms of the cooling time.  Also, and as it should be expected, we obtain larger values for 
the cooling times when considering the pure C core sequences (model F1), since a white dwarf with a 
pure C core cools slower than a white dwarf with a C/O core because of the higher heat capacity of 
C in comparison with that of O, implying a larger amount of energy necessary to change the 
temperature of the core.
 
Some of these white dwarfs have mass estimates from previous investigations.  \cite{sil01} 
calculated masses from gravitational redshifts for WD0315$-$011, WD1354$+$340, WD1544$-$377, 
WD1620$-$391, WD1659$-$531 and WD2253$-$081.  The results of that study are compatible with the 
masses derived in this work except for WD1544$-$377, whose mass is 25\% smaller when calculated 
from its gravitational redshift.  However, \cite{kaw07} inferred the spectroscopic mass of this 
star, together with those of WD1620$-$391 and WD1659$-$531, that are in good agreement with our 
results. WD0913$+$442 was studied by \cite{kar05} and \cite{ber01}. The former inferred the 
spectroscopic mass of the white dwarf and the latter used photometry and the trigonometric parallax 
to estimate the mass.  In both cases, the results are compatible with the value derived here.

\begin{table*}
\begin{center}
\caption{Stellar parameters derived for the observed white dwarfs.}
\begin{tabular}{ccccccc}
\hline
\hline
\noalign{\smallskip}
  & \multicolumn{2}{c}{Model S0$^{\mathrm{a}}$}  & 
    \multicolumn{2}{c}{Model F0$^{\mathrm{b}}$}  & 
    \multicolumn{2}{c}{Model F1$^{\mathrm{b}}$}  \\
\noalign{\smallskip}
\hline
\noalign{\smallskip}
Name & $M_{\rm WD}$ & $t_{\rm cool}$ & $M_{\rm WD}$ & $t_{\rm cool}$ & $M_{\rm WD}$ & $t_{\rm cool}$ \\
     & $(\rm M_{\sun})$ & (Gyr) & $(\rm M_{\sun})$ & (Gyr) & $(\rm M_{\sun})$ & (Gyr) \\
\hline
\noalign{\smallskip}
WD0023$-$109 & $0.56\pm0.03$ & $0.47\pm0.03$   & $0.56\pm0.03$ & $0.50\pm0.03$   & $0.56\pm0.03$ & $0.53\pm0.03$   \\
WD0315$-$011 & $0.60\pm0.20$ & $1.20\pm0.56$   & $0.60\pm0.18$ & $1.28\pm0.45$   & $0.60\pm0.18$ & $1.37\pm0.42$   \\
WD0413$-$077 & $0.54\pm0.02$ & $0.112\pm0.008$ & $0.54\pm0.02$ & $0.11\pm0.01$   & $0.54\pm0.02$ & $0.12\pm0.01$   \\
WD0913$+$442 & $0.78\pm0.01$ & $1.24\pm0.05$   & $0.78\pm0.05$ & $1.24\pm0.15$   & $0.78\pm0.05$ & $1.35\pm0.12$   \\
WD1304$+$227 & $0.73\pm0.02$ & $0.62\pm0.03$   & $0.73\pm0.02$ & $0.68\pm0.03$   & $0.73\pm0.02$ & $0.71\pm0.03$   \\
WD1354$+$340 & $0.50\pm0.04$ & $0.20\pm0.02$   & $0.50\pm0.04$ & $0.19\pm0.02$   & $0.50\pm0.03$ & $0.21\pm0.02$   \\
WD1544$-$377 & $0.78\pm0.02$ & $0.76\pm0.05$   & $0.78\pm0.02$ & $0.81\pm0.04$   & $0.78\pm0.02$ & $0.86\pm0.05$   \\
WD1620$-$391 & $0.63\pm0.01$ & $0.026\pm0.001$ & $0.63\pm0.01$ & $0.022\pm0.001$ & $0.63\pm0.01$ & $0.025\pm0.001$ \\
WD1659$-$531 & $0.66\pm0.01$ & $0.24\pm0.01$   & $0.66\pm0.01$ & $0.25\pm0.01$   & $0.66\pm0.01$ & $0.26\pm0.01$   \\
WD2253$-$081 & $0.75\pm0.09$ & $2.32\pm0.72$   & $0.75\pm0.09$ & $2.20\pm0.44$   & $0.75\pm0.09$ & $2.27\pm0.53$   \\
\noalign{\smallskip}
\hline
\hline
\end{tabular}
\label{tab:5}
\end{center}
\begin{list}{}{}
\item[$^{\mathrm{a}}$] S0 \cite{sal00}
\item[$^{\mathrm{b}}$] F0, F1 \cite{fon01}
\end{list}
\end{table*}

\section{Low-mass companion analysis}

\subsection{Determination of $T_{\rm eff}$.}

\begin{table*}
\begin{center}
\caption{Photometric information available  for the observed FGK stars
         and the effective temperatures derived.}  
\small{
\begin{tabular}{lccccc}
\hline
\hline
\noalign{\smallskip}
Name & $V^1$ & $J$ & $H$ & $K$ & $T_{\rm eff}$ \\
\noalign{\smallskip}
\hline
\noalign{\smallskip}
G158$-$77         & 12.900$\pm$0.020 & 10.740$\pm$0.024 & 10.135$\pm$0.022 & 9.934$\pm$0.019 & $4390\pm30$ \\ 
BD~$-$01~469A$^2$ & 5.370$\pm$0.020  & 3.404$\pm$0.198  & 2.818$\pm$0.210  & 2.678$\pm$0.234 & $4480\pm50$ \\
HD~26965$^3$      & 4.410$\pm$0.020  & 3.013$\pm$0.238  & 2.594$\pm$0.198  & 2.498$\pm$0.236 & $5160\pm35$ \\
BD~$+$44~1847     & 9.000$\pm$0.020  & 7.685$\pm$0.019  & 7.389$\pm$0.017  & 7.315$\pm$0.021 & $5630\pm50$ \\ 
BD~$+$23~2539     & 9.710$\pm$0.020  & 8.458$\pm$0.023  & 8.109$\pm$0.016  & 8.074$\pm$0.017 & $5665\pm50$ \\ 
BD~$+$34~2473     & 9.080$\pm$0.020  & 8.090$\pm$0.030  & 7.851$\pm$0.034  & 7.836$\pm$0.024 & $6270\pm70$ \\ 
HD~140901         & 6.010$\pm$0.020  & 4.959$\pm$0.214  & 4.505$\pm$0.076  & 4.323$\pm$0.016 & $5635\pm50$ \\
HD~147513         & 5.376$\pm$0.020  & 4.405$\pm$0.258  & 4.025$\pm$0.190  & 3.933$\pm$0.036 & $5985\pm75$ \\
HD~153580         & 5.285$\pm$0.020  & 4.453$\pm$0.318  & 4.181$\pm$0.206  & 4.126$\pm$0.036 & $6470\pm90$ \\
BD~$-$08~5980     & 8.030$\pm$0.020  & 6.816$\pm$0.034  & 6.492$\pm$0.033  & 6.353$\pm$0.020 & $5670\pm50$ \\ 
\noalign{\smallskip}
\hline
\hline
\end{tabular}}
\label{tab:ms1}
\end{center}
\small  \footnotemark[1]{Magnitude errors adopted.}\\
\small  \footnotemark[2]{$T_{\rm  eff}$ derived  from  the Eggen  $RI$
                        photometry  using   the  color-$T_{\rm  eff}$
                        relation   from   \cite{hou00}.  $JHK$   2MASS
                        magnitudes were saturated.}\\
\small  \footnotemark[3]{$T_{\rm eff}$  derived from  the  $V-K$ color
                        using  the  color-temperature  relation  from
                        \cite{mas06}.  $JHK$   2MASS  magnitudes  were
                        saturated.}\\
\end{table*}

We have used the available photometry --- $V$ from SIMBAD and $JHK$ from 2MASS (Table 
\ref{tab:ms1})  --- to derive the effective temperatures of these stars, $T_{\rm eff}$, following 
the method of \cite{mas06}.  This procedure consists on calculating synthetic photometry using the 
non-overshoot Kurucz atmosphere model grid (Kurucz 1979)\footnote{{\tt 
http://kurucz.harvard.edu/grids.html}}. Then, we developed a fitting algorithm that is based on the 
minimization of the $\chi^2$ parameter using the Levenberg-Marquardt method.  $\chi^2$ is defined 
from the differences between the observed and synthetic $VHJK$ magnitudes.  This function depends 
indirectly on $T_{\rm eff}$, $\log g$, [Fe/H] and a magnitude difference $\Re$, which is the ratio 
between the synthetic (star's surface)  and the observed flux at Earth, $\Re=-2.5\log(F_{\rm 
star}/F_{\rm earth})$. Tests show that the spectral energy distribution in the optical/IR for the 
range of temperatures that corresponds to FGK stars is only weakly dependent on gravity and 
metallicity, which makes it possible to derive accurate temperatures for stars with poor 
determinations of $\log g$ and [Fe/H]. Taking this into account, we assume initial values of $\log 
g=4.50$ and [Fe/H]=0.0 to estimate the effective temperatures. We did not consider 
interstellar extinction corrections, since they have negligible effects considering the nearby 
distances of the stars under study. The results are given in Table \ref{tab:ms1}.

\begin{table*}
\begin{center}
\caption{$RI$  Eggen  photometry  available  for the  two  stars  with
         saturated 2MASS magnitudes.}  
\footnotesize
\begin{tabular}{lccccccc}
\hline
\hline
\noalign{\smallskip}
Name & $(R-I)_K$ & $(V-R)_K$ & $(V-I)_K$
     & $(R-I)_C$ & $(V-R)_C$ & $(V-I)_C$\\
\noalign{\smallskip}
\hline
\noalign{\smallskip}
BD~$-$01~469A 	& 0.375 & 0.540 & 0.915 & 0.500 & 0.626 & 1.126 \\
HD~26965 	& 0.305 & 0.340 & 0.645 & 0.432 & 0.441 & 0.873 \\ 
\noalign{\smallskip}
\hline
\hline
\end{tabular}
\label{tab:ms2}
\end{center}
\end{table*}

It can be seen from Table \ref{tab:ms1} that the $JHK$ 2MASS magnitudes for BD~$-$01~469A and 
HD~26965 are saturated. Thus, in order to derive accurate effective temperatures for these stars, 
we considered $RI$ photometry (Eggen 1971)  available from The Lausanne Photometric Database (GCPD) 
(Table \ref{tab:ms2}).  We used the relations of \cite{bes79} to transform between Cousins and the 
Kron-Eggen system in order to obtain $(R-I)_C$ and $(V-I)_C$.  Then, we consider the suitable 
color-temperature relations derived by \cite{hou00} to infer their effective temperatures.

In the case of BD~$-$01~469A we obtained $T_{\rm eff}=4525$ K and $T_{\rm eff}=4425$ K, for 
$(V-I)_C$ and $(V-R)_C$, respectively. Using the Stromgren $b-y$ index of $0.633$ also present at 
the same database we obtain $T_{\rm eff}=4500$ K considering the calibration of Olsen (1984). We 
consider as the final result the mean value of these three temperatures, $T_{\rm eff}=4480\pm50$ K.  
The effective temperature obtained for BD~$-$01~469A is 230 K lower than the one reported in 
\cite{mcw90}, who used $(B-V)$ from the Bright Star Catalog (BSC) and the corresponding calibration 
of color-temperature. These authors derived the effective temperature from an extrapolation, since 
their calibration did not cover stars with such low temperatures. Thus, we consider that the value 
that we have obtained is more reliable.

Regarding HD~26965 we used also the relations of \cite{hou00} obtaining $T_{\rm eff}=5200$ K and 
$T_{\rm eff}=5345$ K, for $(V-I)_C$ and $(V-R)_C$ respectively.  Besides the Eggen $RI$ photometry, 
the $J$ and $K$ Johnson magnitudes (Johnson et al.~1968)  of HD~26965 are also available at The 
Lausanne Photometric Database (GCPD).  These photometric data are given in Table \ref{tab:ms3}.  
We used the relation of \cite{bes88} to transform $V-K$ from the Johnson to the Johnson-Glass 
system.  Then, we used the $(V-K)$-temperature calibration of \cite{hou00} obtaining $T_{\rm 
eff}=5135$ K. To compare this result we can use also the $(V-K)$-temperature relation from 
\cite{mas06} that gives $T_{\rm eff}=5185$ K.  We deem the values derived from the $(V-K)$ color 
are more accurate, so, our final value should be the mean of them, $T_{\rm eff}=5160\pm35$ K.  
This value is in reasonable agreement with $T_{\rm eff}=5090$ K, which is the result obtained by 
\cite{ste83} using also the available (Johnson 1966) $V-R$, $V-I$, $V-J$, $V-K$, and $V-L$ colors 
and the \cite{joh66} color calibrations.

\begin{table}
\begin{center}
\caption{Johnson photometry available for HD~26965.}
\small{
\begin{tabular}{lcccccc}
\hline
\hline
\noalign{\smallskip}
Name & $J$ & $H$ & $K$ & $L$ & $(V-K)_{J-G}$ \\
\noalign{\smallskip}
\hline
\noalign{\smallskip}
HD~26965 & 2.95 & 2.48 & 2.41 & 2.37 & 2.005\\
\noalign{\smallskip}
\hline
\hline
\end{tabular}}
\label{tab:ms3}
\end{center}
\end{table}

\begin{table*}[t]
\begin{center}
\caption{Solar  lines chosen to  calibrate the  atomic data  list. The
         $\log gf$'s used by different authors and the ones determined
         in this work are also given, as well as the equivalent widths
         of these lines in the solar spectrum.}  \small{
\begin{tabular}{lccccccccc}
\hline
\hline
\noalign{\smallskip}
Species & Wavelength & LEP  & $\log gf$ & $\log gf$ & $\log gf$  & $\log gf$     & EW${_{\sun}}$ \\
        & (\AA)      & (eV) &   RAL       & RTLA      & AMMSF      & Adopted value & (m\AA)        \\
\noalign{\smallskip}
\hline
\noalign{\smallskip}
Fe~{\sc i}  & 5293.963 & 4.140  & $-1.770$ & $...$    & $...$    & $-1.757$ & 27.3 \\
Fe~{\sc i}  & 5379.574 & 3.690  & $-1.510$ & $...$    & $...$    & $-1.495$ & 52.1 \\
Fe~{\sc i}  & 5386.335 & 4.15 	& $-1.670$ & $...$    & $...$    & $-1.690$ & 31.3 \\
Fe~{\sc i}  & 5543.937 & 4.217  & $-1.040$ & $...$    & $...$    & $-0.980$ & 52.7 \\
Fe~{\sc i}  & 5775.080 & 4.22	& $-1.300$ & $...$    & $-1.155$ & $-1.100$ & 51.9 \\
Fe~{\sc i}  & 5852.217 & 4.549  & $-1.230$ & $-1.170$ & $...$    & $-1.190$ & 38.1 \\
Fe~{\sc i}  & 5856.083 & 4.294	& $-1.460$ & $-1.560$ & $...$    & $-1.520$ & 36.7 \\
Fe~{\sc i}  & 5859.600 & 4.550  & $...$    & $-0.610$ & $...$    & $-0.450$ & 60.1 \\
Fe~{\sc i}  & 6027.050 & 4.076  & $-1.090$ & $-1.170$ & $...$    & $-1.080$ & 58.9 \\
Fe~{\sc i}  & 6078.999 & 4.652	& $-1.020$ & $...$    & $-1.123$ & $-0.930$ & 46.8 \\ 
Fe~{\sc i}  & 6151.617 & 2.176  & $-3.300$ & $-3.280$ & $-3.486$ & $-3.290$ & 53.3 \\
Fe~{\sc i}  & 6165.360 & 4.143  & $-1.460$ & $-1.460$ & $-1.645$ & $-1.440$ & 47.7 \\
Fe~{\sc i}  & 6170.504 & 4.765  & $-0.380$ & $...$    & $...$    & $-0.200$ & 63.2 \\
Fe~{\sc i}  & 6173.341 & 2.223	& $-2.880$ & $-2.880$ & $...$    & $-2.840$ & 66.9 \\
Fe~{\sc i}  & 6200.314 & 2.609	& $-2.440$ & $-2.440$ & $...$    & $-2.330$ & 66.7 \\
Fe~{\sc i}  & 6322.694 & 2.588  & $-2.430$ & $-2.430$ & $-2.503$ & $-2.395$ & 68.3 \\
Fe~{\sc i}  & 6481.869 & 2.279  & $-2.980$ & $-2.970$ & $...$    & $-2.940$ & 65.9 \\
Fe~{\sc i}  & 6713.771 & 4.796  & $...$    & $-1.390$ & $-1.606$ & $-1.425$ & 35.5 \\
Fe~{\sc i}  & 6857.243 & 4.076  & $...$    & $-2.040$ & $-2.203$ & $-2.055$ & 23.4 \\
Fe~{\sc i}  & 7306.556 & 4.178  & $...$    & $...$    & $-1.684$ & $-1.545$ & 51.7 \\
Fe~{\sc i}  & 7802.473 & 5.086  & $...$    & $-1.310$ & $-1.493$ & $-1.350$ & 21.2 \\
Fe~{\sc i}  & 7807.952 & 4.990  & $...$    & $-0.510$ & $-0.602$ & $-0.477$ & 66.0 \\
Fe~{\sc ii} & 5197.577 & 3.230  & $-2.220$ & $...$	 & $...$    & $-2.330$ & 58.2 \\
Fe~{\sc ii} & 5234.625 & 3.221  & $-2.180$ & $-2.220$ & $...$    & $-2.285$ & 57.6 \\
Fe~{\sc ii} & 6149.258 & 3.889  & $...$    & $-2.630$ & $-2.858$ & $-2.770$ & 34.6 \\
Fe~{\sc ii} & 6247.560 & 3.892  & $...$    & $-2.270$ & $-2.770$ & $-2.385$ & 47.6 \\
Fe~{\sc ii} & 6369.460 & 2.891  & $...$    & $-4.020$ & $...$    & $-4.190$ & 26.2 \\
Fe~{\sc ii} & 6456.383 & 3.903  & $...$    & $-2.060$ & $-2.209$ & $-2.145$ & 56.1 \\
Fe~{\sc ii} & 6516.081 & 2.891  & $-3.310$ & $...$    & $...$    & $-3.415$ & 49.3 \\
Si~{\sc i}  & 5948.540 & 5.082  & $-1.130$ & $...$    & $-1.098$ & $-1.170$ & 85.7 \\
Si~{\sc i}  & 6721.848 & 5.863  & $...$    & $-1.060$ & $-1.100$ & $-1.090$ & 45.0 \\
Si~{\sc ii} & 6371.360 & 8.120  & $...$    & $-0.050$ & $...$    & $-0.080$ & 45.2 \\
Si~{\sc ii} & 6800.596 & $...$  & $...$    & $...$    & $...$    & $-1.715$ & 14.9 \\
Ni~{\sc i}  & 5805.213 & 4.168  & $...$    & $...$    & $-0.530$ & $-0.570$ & 40.5 \\
Ni~{\sc i}  & 6176.820 & 4.088  & $...$    & $-0.260$ & $-0.148$ & $-0.118$ & 59.1 \\
Ni~{\sc i}  & 6378.260 & 4.154  & $...$    & $-0.830$ & $...$    & $-0.820$ & 34.4 \\
Ni~{\sc i}  & 6772.320 & 3.658  & $...$    & $-0.970$ & $...$    & $-0.900$ & 55.0 \\
Ni~{\sc i}  & 7555.598 & 3.848  & $...$    & $...$    & $ 0.069$ & $ 0.059$ & 85.7 \\
\noalign{\smallskip}			
\hline					
\hline					
\end{tabular}				
\label{tab:ms4}}			
\end{center}				
References. (RAL) Ram{\'{\i}}rez et al.~2007; (RTLA) Reddy et al.~2003; (AMMSF) Affer et al.~2005\\
\end{table*}

\subsection{Determination of [Fe/H]}

To derive the metallicity of the stars we fitted the observed absorption lines with synthetic 
spectra computed with SYNSPEC (Hubeny \& Lanz 1995)\footnote{{\tt 
http://nova.astro.umd.edu/Synspec43/synspec.html}} and Kurucz's model atmospheres (Kurucz 1993).  
For each star, we used the model corresponding to the derived $T_{\rm eff}$ and assumed a value for 
$\log g$.  SYNSPEC is a program for calculating the spectrum emergent from a given model 
atmosphere. SYNSPEC was originally designed to synthesize spectra from atmospheres calculated using 
TLUSTY (Lanz \& Hubeny 1995), but may also be used with other model atmospheres as input (e.g. LTE 
Kurucz's ATLAS models, as in our case).  The program is complemented by the routine ROTINS that 
calculates the rotational and instrumental convolutions for the net spectrum produced by SYNSPEC.

Line selection and atomic data calibration is a crucial step to derive the metallicity of a star.  
We selected the lines from two sources: \cite{red03} and \cite{ram07} taking into account different 
requirements. The suitable stellar lines should have a relatively small equivalent width, i.e., 
$\Delta W_\lambda <50 \;{\rm m}\AA$ approximately. We discarded also the lines which fell in the 
spectral gaps between the spectral orders or those that appeared asymmetric, which were assumed to 
be blended with unidentified lines. It is very important also to consider lines for the same 
species but corresponding to different transitions and ionization states, since this can provide 
useful cross-checks to test if the derived effective temperature is correct.  This is particularly 
interesting when stars are cooler, since it is more difficult to derive the temperature with 
accuracy.  We selected also some stellar lines farther in the red part of the spectrum from the 
linelist of \cite{aff05}.

\begin{figure}[t]
\begin{center}
\includegraphics[clip,width=0.9\columnwidth]{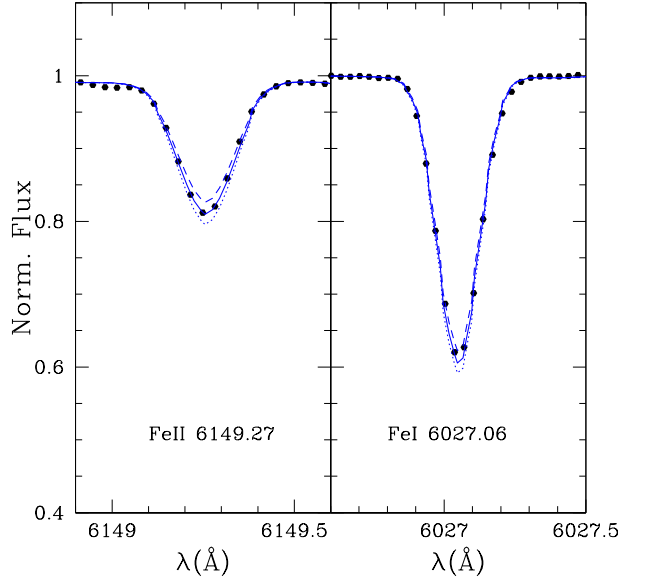}
\caption{Fits of the observed spectra for HD 140901. The solid line is
         the fit corresponding  to the derived $Z$ and  the dotted and
         dashed  lines are spectra  computed for  $+1 \sigma$  and $-1
         \sigma$ from the average.}
\label{fig:fitms}
\end{center}
\end{figure}

The first step of this procedure is to calibrate the atomic data list using the Kurucz's solar 
spectrum\footnote{{\tt http://kurucz.harvard.edu/sun.html}} and the corresponding solar atmosphere, 
which has $T_{\rm eff}=5777$ K, $\log g=4.437$ and $\xi=1.5\; {\rm km\;  s^{-1}}$. For each 
selected line we changed the oscillator strength ($\log gf$) in the Kurucz's atomic linelist until 
it reproduced the observed solar spectrum.  In Table \ref{tab:ms4} we give the values of \cite{ram07}, \cite{red03} and \cite{aff05}, and the adopted values 
that we have used in our analysis. The equivalent widths of the fitted solar lines measured with 
the IRAF task {\tt splot} are given as well.  Therefore, the oscillator strengths will be fixed 
when fitting the spectra of the FGK stars that we have observed. After discerning which lines were 
suitable for the fitting procedure we selected the value of $\log g$ (3.5, 4.0 or 4.5) that gave 
the same abundances for different species and different ionization states. We estimated the 
value of the microturbulence, $\xi$, using the relationship derived by \cite{all04} as function of 
$T_{\rm eff}$ and $\log g$ --- see Table \ref{tab:ms5}). After obtaining the metallicity 
considering the proper $\log g$ and $\xi$, we recalculated the $T_{\rm eff}$ performing again the 
corresponding fit to synthetic photometry, which led to negligible adjustments. Another 
parameter that could affect the determination of metallicity is the macroturbulence. We adjusted this 
parameter using a rotational profile and a Gaussian broadening function independently. Both 
approximations led to the same metallicities. In 
Fig.~\ref{fig:fitms} we show the spectral fits for one of the companions of the DA white dwarfs 
(HD~140901).  We have chosen to plot the fits corresponding to Fe~{\sc i} and Fe~{\sc ii}, to show 
how the method works for different ionization states.

\subsection{Age determination}

\begin{table*}
\begin{center}
\caption{Stellar parameters derived for the observed FGK stars.}
\small{
\begin{tabular}{lccccccc}
\hline
\hline
\noalign{\smallskip}
Name & $  T_{\rm eff}$ &  $\xi$ & [Fe/H] & $Z$ & $\log g$ & $\log(L/L_{\sun})$ & Isoch. Age \\
 & (K) & (${\rm km\; s^{-1}}$) & & & (dex) & & (Gyr) \\
\noalign{\smallskip}
\hline
\noalign{\smallskip}
G158$-$77$^{1,2}$   & $4390\pm30$ & $...$   & $...$           & $...$           & $...$ & $...$            & $...$                  \\ 
BD~$-$01~469A$^{3}$ & $4480\pm50$ & $0.898$ & $-0.10\pm0.08$  & $0.016\pm0.003$ & $3.5$ & $1.669\pm0.109$  & $4.17^{+3.04}_{-2.05}$ \\
HD~26965            & $5140\pm15$ & $1.137$ & $-0.41\pm0.07$  & $0.008\pm0.001$ & $4.5$ & $-0.344\pm0.094$ & $...$                  \\
BD~$+$44~1847       & $5630\pm50$ & $1.305$ & $-0.44\pm0.05$  & $0.007\pm0.001$ & $4.5$ & $-0.184\pm0.059$ & $...$                  \\ 
BD~$+$23~2539$^{1}$ & $5665\pm50$ & $1.317$ & $0.03\pm0.06$   & $0.021\pm0.003$ & $4.5$ & $...$            & $...$                  \\ 
BD~$+$34~2473       & $6270\pm70$ & $1.524$ & $-0.12\pm0.04$  & $0.015\pm0.002$ & $4.5$ & $0.379\pm0.109$  & $3.26^{+0.74}_{-1.46}$ \\ 
HD~140901           & $5635\pm50$ & $1.306$ & $0.02\pm0.07$   & $0.021\pm0.003$ & $4.5$ & $-0.055\pm0.012$ & $...$                  \\
HD~147513           & $5985\pm75$ & $1.426$ & $0.001\pm0.077$ & $0.020\pm0.003$ & $4.5$ & $0.029\pm0.017$  & $...$                  \\
HD~153580$^{4}$     & $6470\pm90$ & $1.592$ & $-0.01\pm0.01$  & $0.019\pm0.004$ & $4.5$ & $0.671\pm0.042$  & $2.51^{+0.34}_{-0.32}$ \\
BD~$-$08~5980	    & $5670\pm50$ & $1.318$ & $-0.37\pm0.05$  & $0.008\pm0.001$ & $4.5$ & $-0.140\pm0.040$ & $...$                  \\
\noalign{\smallskip}
\hline
\hline
\end{tabular}
\label{tab:ms5}}
\end{center}
\small \footnotemark[1]{The distance of these stars is not known.}\\  
\small \footnotemark[2]{Low S/N.}\\  
\small \footnotemark[3]{[Fe/H] taken from \cite{mcw90}.}\\  
\small \footnotemark[4]{[Fe/H] taken from \cite{tay03}.}\\  
\end{table*}

\begin{table*}
\caption{X-ray  parameters given by  the ROSAT  Catalog and  the ages derived.}
\begin{center}
\footnotesize
\begin{tabular}{lccccc}
\hline
\hline
\noalign{\smallskip}
Name & $HR$  & Count Rate  & $\log(L_x)$ & Age \\
 & & (c/s) &  & (Gyr) \\
\noalign{\smallskip}
\hline
\noalign{\smallskip}
HD~26965  & $-0.28\pm0.06$ & $0.796\pm0.052$ & $28.22\pm0.12$ & $1.07\pm0.37$\\
HD~140901 & $-0.73\pm0.11$ & $0.150\pm0.023$ & $28.27\pm0.31$ & $0.94\pm0.50$\\
HD~147513 & $-0.25\pm0.06$ & $0.650\pm0.045$ & $28.95\pm0.14$ & $0.33\pm0.12$\\
\noalign{\smallskip}
\hline
\hline
\end{tabular}
\label{tab:ms6}
\end{center}
\end{table*}

\begin{figure}[!t]
\centering
\includegraphics[clip,width=0.9\columnwidth]{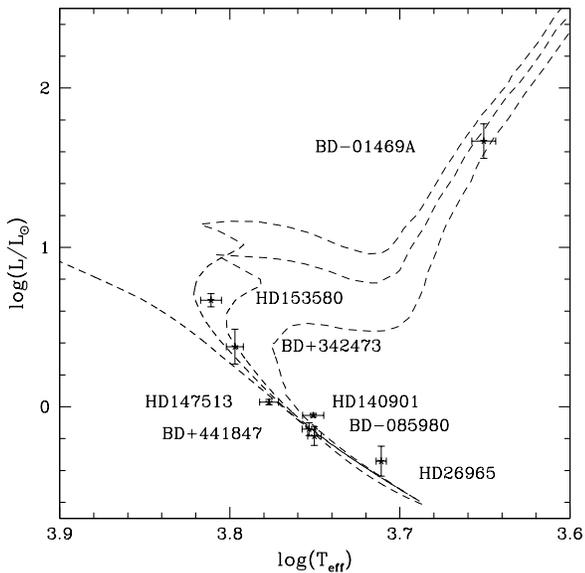}
\caption{Hertzsprung-Russell   diagram   for   the  companions.    The
         isochrones of \cite{sch92} for different ages (ZAMS, 2, 3 and
         7 Gyr,  from left  to right) and  solar metallicity  are also
         plotted.}
\label{fig:lumteff}
\end{figure}

\begin{figure}[!t]
\centering
\includegraphics[clip,width=0.9\columnwidth]{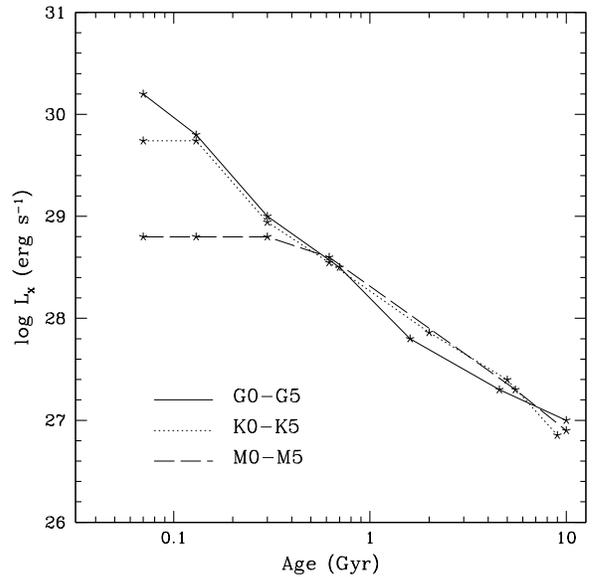}
\caption{X-ray luminosity versus age for stars with different spectral
         types according to \cite{rib07}.}
\label{fig:xlum}
\end{figure}

For most of our stars in our sample the parallax is known (from the Hipparcos Catalogue), thus, the calculation 
of the luminosity, $L$, is straightforward using the apparent magnitude after estimating the 
bolometric magnitude, $M_{\rm bol}$. For best accuracy we have used the $K$ band magnitude and the 
bolometric corrections of \cite{mas06}.  In Fig.~\ref{fig:lumteff} we show the Hertzsprung-Russell 
diagram for the FGK stars in our list with known distances.  The isochrones of \cite{sch92} for 
different ages and solar metallicity have been also plotted to show at which evolutionary state 
these stars are. As can be seen, the isochrone fitting technique is suitable for BD~$+$34~2473, 
HD~153580 (both F stars) and for BD~$-$01~469A (K subgiant).  The rest of stars are too close to 
the ZAMS and hence the use of isochrones does not provide accurate values for their ages. When the 
isochrone fitting is appropriate, we have performed an interpolation in the grid of stellar models 
of \cite{sch92} considering the derived $T_{\rm eff}$, $Z$ and $L$ to obtain the ages of these 
stars, i.e., the total ages of the white dwarfs in the common proper motion pairs.  Our results are 
given in Table \ref{tab:ms5}.

Another age indicator which could be used is X-ray luminosity. For some of these objects there are 
data available from the ROSAT All-Sky Bright Source Catalogue --- 1RXS (Voges 1999)  --- which 
gives the count rate (number of detected counts per second) and the hardness ratio, $HR$.  The 
hardness ratio is defined $HR=(H-S)/(H+S)$, where $H$ and $S$ are respectively the counts recorded 
in the hard and soft PSPC pulse height channels.  To obtain the X-ray flux of a given star, we 
considered the calibrations of \cite{sch95}. In particular, we used the conversion factor to 
obtain the energy flux from the measured count rate, which depends on $HR$:

\begin{equation}
CF=(5.30HR+8.31)\times10^{-12}\;{\rm ergs}\;{\rm cm}^{-2}\;{\rm counts}^{-1}
\end{equation}

\cite{rib07} calculated a relationship between the age and X-ray luminosity for stars of different 
spectral types (Fig.~\ref{fig:xlum}) using both cluster data and stars belonging to wide binaries, 
or using kinematic criteria.  In Table \ref{tab:ms6} we give the ROSAT information regarding these 
objects, the X-ray luminosity and the ages derived for the FGK companions with X-ray emission. 
The errors of the ages have been calculated considering the errors in the X-ray luminosity and an 
assumed cosmic dispersion for each relation (8 and 20\% for G and K stars, respectively). There is 
also ROSAT information available for HD~153580, but since it is a member of a spectroscopic binary 
these relations cannot be applied.

\begin{table*}[t]
\caption{Ages, metallicities and masses  for the white dwarfs in these
         common proper motion pairs.}
\begin{center}
\begin{tabular}{lccccccc}
\hline
\hline
\noalign{\smallskip}
WD &  Age & $t_{\rm cool}$ & $t_{\rm MS}$ & $M_{\rm F}$ & $M_{\rm I}$  & $Z$ \\
   &  (Gyr) & (Gyr)        & (Gyr)        & ($\rm M_{\sun}$) & ($\rm M_{\sun}$) & \\
\noalign{\smallskip}
\hline
\noalign{\smallskip}
WD0315$-$011 & $4.17^{+3.04}_{-2.05}$ & $1.20\pm0.56$   & $2.97^{+3.09}_{-2.12}$ & $0.60\pm0.20$ & $1.48^{+0.87}_{-0.28}$ & $0.016\pm0.003$ \\
WD0413$-$017 & $1.07\pm0.37$          & $0.112\pm0.008$ & $0.96\pm0.37$          & $0.54\pm0.02$ & $2.07^{+0.53}_{-0.27}$ & $0.008\pm0.001$\\
WD1354$+$340 & $3.26^{+0.74}_{-1.46}$ & $0.20\pm0.02$   & $3.06^{+0.74}_{-1.46}$ & $0.50\pm0.04$ & $1.46^{+0.31}_{-0.09}$ & $0.015\pm0.002$\\ 
WD1544$-$377 & $0.94\pm0.50$          & $0.76\pm0.05$   & $0.18\pm0.50$          & $0.78\pm0.02$ & $4.13^{+?}_{-1.49}$ & $0.021\pm0.003$ \\
WD1620$-$391 & $0.33\pm0.12$          & $0.026\pm0.001$ & $0.30\pm0.12$         & $0.63\pm0.01$ & $3.45^{+0.65}_{-0.35}$ & $0.020\pm0.003$ \\
WD1659$-$531 & $2.51^{+0.34}_{-0.32}$ & $0.24\pm0.01$   & $2.27^{+0.34}_{-0.32}$ & $0.66\pm0.01$ & $1.58^{+0.08}_{-0.05}$ & $0.019\pm0.004$ \\
\noalign{\smallskip}
\hline
\hline
\end{tabular}
\label{tab:mif}
\end{center}
\end{table*}

\section{The initial-final mass relationship}

\begin{figure}[h]
\begin{center}
\includegraphics[scale=0.40]{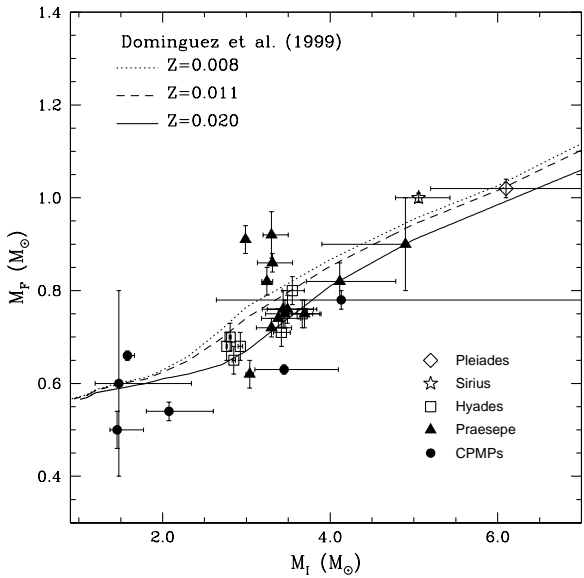}
\caption{Final  masses versus  initial  masses for  the common  proper
         motion pairs studied here and some precise available data.}
\label{fig:mif}
\end{center}
\end{figure}

Once we know the total age of the white dwarfs and the metallicity of their progenitors, the 
initial masses can be derived considering suitable stellar models.  In our case we have used the 
stellar tracks of \cite{dom99}. The initial and final masses obtained are detailed in Table 
\ref{tab:mif}.  Other parameters, such as overall ages, cooling times, main-sequence lifetimes of 
the progenitors and metallicities are also given. As can be noted all the total ages exceed the 
cooling times, as expected.

In Fig.~\ref{fig:mif} we represent the final masses versus the initial masses obtained for the
white dwarfs in our sample for which the age and metallicity have been derived. The lines correspond 
to the theoretical initial-final mass relationships of \cite{dom99} for 
different metallicities.  For the sake of comparison we have also included the most precise data 
that are currently being used to define the semi-empirical initial-final mass relationship.  For 
the Hyades and Praesepe, we plot the results obtained by \cite{cla01}, and some recent results from 
Dobbie et al.~(2004, 2006). We also used the results of \cite{dob06} for the only known Pleiades 
white dwarf. In the case of Sirius, we have used the initial and final masses derived by 
\cite{lie05b}.

From an inspection of Fig.~\ref{fig:mif} it can be noted that the observational data present large 
dispersion, which is higher than the uncertainties, in comparison with the theoretical 
initial-final mass relationships of \cite{dom99}.  According to our results, a main-sequence star of 
$1.5\, \rm M_{\sun}$ with approximately solar metallicity could end up as white dwarfs with masses 
that differ by $\sim25\%$ (cf.~WD1354$+$340 and WD1659$-$531).  Moreover, two white dwarfs of 
nearly the same masses could come from main-sequence stars with masses different by a factor of 2 
(cf.~WD1620$-$391 and WD1659$-$531).  Apparently, this difference is not a consequence of 
metallicity, since it is practically the same for these objects (Table \ref{tab:mif}).  However, it 
is also interesting to note that the influence of metallicity on the theoretical initial-final mass 
relationship seems to be almost negligible below $2\, \rm M_{\sun}$. Other factors, such as 
magnetic fields or rotation (Dom{\'{\i}}nguez et al.~1996) should be studied in detail in order to 
discern their effect on this relation.

The ages of star clusters are usually calculated to a higher accuracy than in the case of the 
individual low-mass stars considered in this work, which should allow to obtain the initial masses with better 
accuracy.  However, from Fig.~\ref{fig:mif} it can be noted that white dwarfs in clusters display a 
large dispersion, especially between $3$ and $4\, \rm M_{\sun}$. Thus, this 
scatter in the observational data seems to be a real effect, rather than a consequence of the 
uncertainties in the mass estimates. Hence, there is no apparent reason for which the initial-final 
mass relationship should be considered a single-valued function.  A thorough complete comparison of 
our results based in common proper motion pairs with cluster data will be discussed in a 
forthcoming paper (Catal\'an et al.~2008).

One of the most important contributions of our work is the study of the range of initial masses 
corresponding to $1.5-2\, \rm M_{\sun}$, which was not covered by the research based on open 
cluster data (Ferrario et al.~2005, Dobbie et al.~2006). The recent study of \cite{kal07} 
using old open clusters has also provided some new data in the low-mass domain. It is worth to 
mention that 5 of the 6 white dwarfs of our final sample have masses near the typical values 
derived by, e.g., Kepler et al.~(2007), $M\sim0.6\,\rm M_{\sun}$, which represent 90\% of the white 
dwarfs found in the SDSS. This stems from the fact that the progenitors of white dwarfs in open 
clusters were usually more massive ($M>2\,\rm M_{\sun}$) since clusters are relatively young and 
the low-mass stars, which would produce the typical white dwarfs, are still on the main sequence.  
Since some of the pairs that we have studied have larger ages than the typical values for open 
clusters, the white dwarfs that belong to these pairs can be less massive. Thus, we consider that white dwarfs in 
common proper motion pairs are more representative of the Galactic white dwarf field population 
than white dwarfs in open clusters.

\section{Summary and Conclusions}

We have studied a sample of common proper motion pairs comprised of a white dwarf and a FGK star.  
We have performed high signal-to-noise low resolution spectroscopy of the white dwarf members, 
which led us to carry out a full analysis of their spectra and to make a re-classification when 
necessary.  From the fit of their spectra to white dwarf models we have derived their atmospheric 
parameters. Then, using different cooling sequences --- namely those of \cite{sal00} and 
\cite{fon01} --- their masses and cooling times were obtained.  Simultaneously, we have performed 
independent high resolution spectroscopic observations of their companions.  Using the available 
photometry we have obtained their effective temperatures.  Then, from a detailed analysis of their 
spectra and using either isochrones or X-ray luminosities, we have derived their metallicities and 
ages (i.e., the metallicities of the progenitors of the white dwarfs and their total ages).

These observations allowed us to obtain the initial and final masses of six white dwarfs in common 
proper motion pairs, four of them corresponding to initial masses below $2\,\rm M_{\sun}$, a range 
which has not been previously covered by the open cluster data.  Our semi-empirical relation shows 
significant scatter, compatible with the results obtained by \cite{fer05} and \cite{dob06}, which 
are mainly based on open cluster data.  However, the dispersion of the results is higher than the 
error bars, which leaves some open questions that should be studied in detail (e.g., rotation or 
magnetic fields).

We have shown that common proper motion pairs containing white dwarfs can be useful to improve 
the initial-final mass relationship, since they cover a wide range of ages, masses and 
metallicities, and they are also representative of the disk white dwarf population.  We have seen 
that the accuracy in the total ages depends almost exclusively on the evolutionary state of the 
low-mass companions. Such relative accuracy becomes poor when the star is close to the ZAMS. 
However, this limitation may not be critical to many common proper motion pairs. Planned deep 
surveys like GAIA, LSST or the Alhambra Survey will discover thousands of new white dwarfs, some of 
them belonging to wide binaries. In the meantime, our most immediate priority is to further extend 
the sample of wide binaries valid for this study.  We are working in the search for more wide 
binaries of our interest in the NLTT catalog (Gould \& Chanam\'e 2004) and also in the LSPM-north 
catalog (L\'epine \& Bongiorno 2007). Detailed study of the current and future common proper motion 
pairs of this type should help to explain the scatter in the semi-empirical initial-final mass 
relationship and to discern whether this is a single-valued function. If consistency between 
observations and theoretical calculations is found, this would have a strong impact on stellar 
astrophysics, since this relationship is used in many different areas, such as chemical evolution 
of galaxies, the determination of supernova rates or star formation and feedback processes in 
galaxies.

\begin{acknowledgements}
We thank D.~Koester for his useful comments in the fitting procedure and for providing us with his 
white dwarf models.  We also wish to thank P.~Bergeron for kindly sharing his fitting routines that 
were very useful to compare with our results.  Finally, we are grateful to T.~Oswalt \& M.~Rudkin 
for the observations of WD0315$-$011. S.~C. would like to acknowledge support from MEC through a 
FPU grant. C.~A.~P ackowledges support from NASA (NAG5-13057, NAG5-13147). This research was 
supported in part by the MEC grants AYA05--08013--C03--01 and 02, by the European Union FEDER funds 
and by the AGAUR. 
\end{acknowledgements}

\end{document}